\documentclass[%
 reprint,
superscriptaddress,
 amsmath,amssymb,
 aps,
 prl,
]{revtex4-2}

\usepackage[version=4]{mhchem}
\usepackage{graphicx}
\usepackage{dcolumn}
\usepackage{bm}

\usepackage{textcomp}

\makeatletter
\def\maketitle{
\@author@finish
\title@column\titleblock@produce
\suppressfloats[t]}
\makeatother

\begin{document}

\preprint{APS/123-QED}

\title{Nanosecond electron imaging of transient electric fields and material response}


\author{Thomas E Gage}
\thanks{These two authors contributed equally to this work.}
\affiliation{Center for Nanoscale Materials, Argonne National Laboratory, Lemont, Illinois 60439, USA}%
\author{Daniel B Durham}%
\thanks{These two authors contributed equally to this work.}
\affiliation{Materials Science Division, Argonne National Laboratory, Lemont, Illinois 60439, USA}%

\author{Haihua Liu}
\affiliation{Center for Nanoscale Materials, Argonne National Laboratory, Lemont, Illinois 60439, USA}%

\author{Supratik Guha}
\affiliation{Materials Science Division, Argonne National Laboratory, Lemont, Illinois 60439, USA}%
\affiliation{Pritzker School for Molecular Engineering, The University of Chicago, Chicago, Illinois 60637, USA}

\author{Ilke Arslan}
\affiliation{Center for Nanoscale Materials, Argonne National Laboratory, Lemont, Illinois 60439, USA}%

\author{Charudatta Phatak}
\email{cd@anl.gov}
\affiliation{Materials Science Division, Argonne National Laboratory, Lemont, Illinois 60439, USA}%

\date{\today}

\begin{abstract}
Electrical pulse stimulation drives many important physical phenomena in condensed matter as well as in electronic systems and devices. Often, nanoscopic and mesoscopic mechanisms are hypothesized, but methods to image electrically driven dynamics on both their native length and time scales have so far been largely undeveloped. Here, we present an ultrafast electron microscopy approach that uses electrical pulses to induce dynamics and records both the local time-resolved electric field and corresponding material behavior with nanometer-nanosecond spatiotemporal resolution. Quantitative measurement of the time-dependent field via the electron beam deflection is demonstrated by recording the field between two electrodes with single-ns temporal resolution. We then show that this can be applied in a material by correlating applied field with resulting dynamics in \ce{TaS2}. First, time-resolved electron diffraction is used to simultaneously record the electric field and crystal structure change in a selected region during a 20 ns voltage pulse, showing how a charge density wave transition evolves during and after the applied field. Then, time-resolved nanoimaging is demonstrated, revealing heterogeneous distortions that occur in the freestanding flake during a longer, lower amplitude pulse. Altogether, these results pave the way for future experiments that will uncover the nanoscale dynamics underlying electrically driven phenomena.
\end{abstract}

\maketitle

\section{\label{sec:intro}Introduction}

Electrical transients induce important and fascinating physics across many disciplines. In condensed matter, for instance, they can cause phase transitions\cite{chae2005vo2,vaskivskyi2016fast,legallo2020pcm}, dielectric breakdown\cite{budenstein1980db,wu2004nspulsedb,lombardo2005dielectric,sasaki2022hbndb}, magnetization switching\cite{yang2017ultrafast,jhuria2020sot}, polarization switching\cite{li2004polswitchfe,chen2015thzbfo,aabrar2022beol}, and electrochemical processes\cite{bagdzevicius2017interface,talin2022ecram}. They can also generate electrodynamics in radiofrequency electronics and actuation in microelectromechanical systems\cite{yao2000rf,varadan2003rf}. While global property measurements offer much insight, the underlying processes can span length scales from local atomic arrangements to nanoscopic interfaces and defects to mesoscopic domains, filaments, geometry, and other variations. Observing the dynamics at these varying length scales is important for a holistic understanding of the mechanisms behind electrically driven phenomena.

While many pump-probe nanoimaging techniques have been developed with the required spatiotemporal resolution, including ultrafast x-ray\cite{mesler2007uxrm,wessels2014uxrm,wen2019uxrm}, electron\cite{zewail2010UEM,lagrange2008dtem}, and near-field optical nanoimaging\cite{stark1995NSOM,nechay1999femtosecond}, they have so far mainly used optical excitation. 
Among these, ultrafast electron probes are especially promising for studying electrically driven dynamics as electrons are sensitive to electric fields\cite{haas2019holodpc} in addition to magnetic fields as well as material properties such as crystal structure and chemical composition. There have been a few recent efforts to apply ultrafast electron characterization to electrically stimulated dynamics. Ultrafast electron diffraction was used to examine phase transition dynamics in \ce{VO2} during electrical pulses, but rise time of delivered excitation was limited to microseconds and real-space imaging was not accessible\cite{SLAC2021ElectricalUED}. Elsewhere, real-space imaging of free-space electrodynamics in a microstrip device during continuous radiofrequency wave excitation was achieved with nanometer-picosecond resolution\cite{Zhu2020MicrowaveUEM}. However, an ability to correlate electrical pulse stimulation and resulting material dynamics across length scales is still a new and important frontier.

Here, we present an ultrafast transmission electron microscopy (TEM) approach that enables correlative measurement of the local electric field, atomic structure change, and nanoscopic real-space dynamics in matter during discrete electrical pulses down to nanosecond timescales. We first demonstrate accurate measurement of the time-resolved field in a free-space region between two electrodes. We then correlate the time-resolved field with induced structural phase transition and emergence of heterogeneous strain in a freestanding \ce{TaS2} flake by utilizing diffraction and imaging modes.

\begin{figure*}[ht]
    \centering
    \includegraphics[width=6.8in]{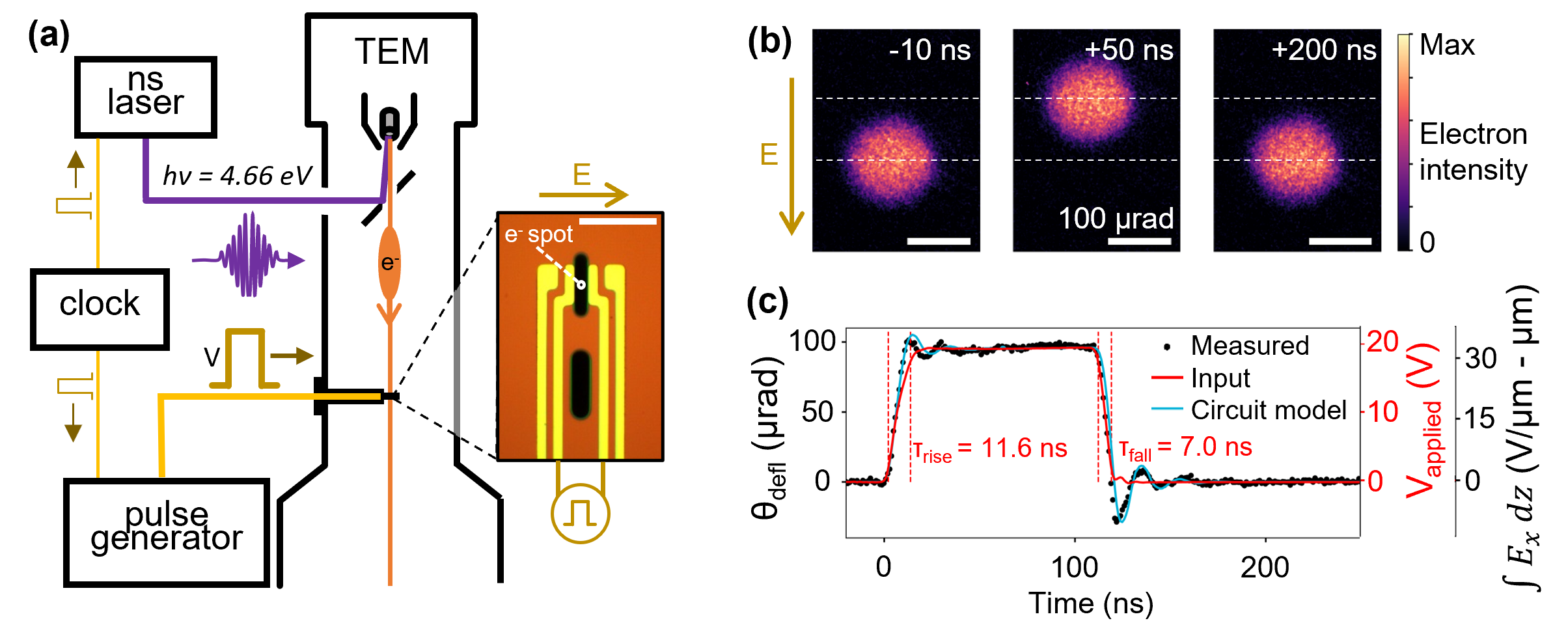}
    \caption{Electrical ultrafast electron microscopy. a) Schematic of the setup. A delay generator (``clock'') outputs trigger signals to a pulse generator and ns laser to generate synchronized pump and probe pulses with a set delay time. UV laser pulses are sent to the photocathode in the TEM to generate sub-ns electron probe pulses, while voltage pulses are sent to stimulate dynamics in the sample. Inset shows the biasing chip used in this work (scale bar = 25 \textmu m), with the position of the electron beam for the following deflection measurements labeled. The reddish-brown \ce{SiNx} membrane supports the gold electrodes, with through holes appearing black. Voltage is applied across the inner two electrodes, giving an electric field along the arrow indicated.  b) Electron beam images in the Fourier plane recorded before, during, and after a 100 ns, 20 V pulse. Scale bar is 100 \textmu rad. Corresponding electric field direction indicated by the arrow: note that these images are rotated by 90 degrees with respect to the optical microscope image. c) Time-resolved beam deflection during the pulse. The measurement (black) is fit with the input pulse shape measured using a GHz oscilloscope (red) and the circuit model result (cyan) with amplitude and starting time as the only fitting parameters.} 
    \label{fig:defl}
\end{figure*}

\section{\label{sec:UEMsetup}Electrical ultrafast electron microscopy setup}
The experiments were conducted using the ultrafast electron microscope (UEM) located at the Center for Nanoscale Materials at Argonne National Laboratory\cite{liu2021anluem}. This specialized microscope is a modified JEOL JEM 2100Plus TEM with optical paths that enable access to the electron gun and specimen. In this study, we utilized the optical access to the gun to generate photoelectrons from a guard ring cathode (AP Tech) with a diameter of 50 \textmu m LaB\textsubscript6 surrounded by 500 \textmu m diameter graphite. Additionally, the microscope includes an extra C\textsubscript0 lens that enhances the electron throughput efficiency. 

The configuration for electrical pump - electron probe experiments developed for this work is shown in Figure \ref{fig:defl}a. Sub-ns photoelectron pulses are produced using a Bright Solutions Wedge Q-switched DPSS laser, which emits at 266 nm and has a pulse duration of 420 ps. The laser can operate from single shot to 100 kHz, and we used 10 kHz and 80 nJ pulses for this study. Meanwhile, we employed two Berkeley Nucleonics 575 delay/pulse generators to regulate the timing of the system and generate electrical pump pulses. Computer software controls the timing of the trigger pulse sent to the Q-switched laser, enabling changing the delay between the arrival of the electron probe pulses and electrical excitation. Our system can generate square wave electrical excitation pulses with up to 20 V amplitude at the BNC output, which was utilized to excite the specimen via a DENSsolutions Lightning heating and biasing holder.

\section{\label{sec:TransientField}Recording nanosecond voltage pulses}

We first demonstrate local field measurement with nanosecond time resolution by recording the deflection of the electron beam. A commercial in-situ heat/bias chip was installed in the TEM holder which includes gold electrodes separated by a vacuum gap at the end, shown in Figure~\ref{fig:defl}a. The probed region was selected using an aperture with an effective 1.2 \textmu m diameter in the center of the 5 \textmu m gap. The electron optics were configured for diffraction-space (Fourier) imaging at 200 cm camera length to record the angular deflection of the beam. Voltage pulses with 20 V amplitude and 100 ns duration were delivered via the inner two electrodes at 10 kHz repetition rate, and Fourier images were recorded for varying pump-probe delay using 5 second exposures. Two time series with 1 ns steps were recorded: one by sequentially increasing the delay, and the other by decreasing. The time-dependent center of mass shift along the field direction was calculated for both measurements\cite{SI} and then, after removing slow drift components by fitting a parabolic background, they were averaged together to give the result shown in Figure~\ref{fig:defl}c. 

Firstly, a distinct deflection due to the electrical pulse can be seen by comparing the Fourier images recorded before, during, and after the pulse, shown in Figure~\ref{fig:defl}b. Moreover, as shown in Figure~\ref{fig:defl}c, the measured time-dependent deflection well matches the input voltage pulse shape, as separately measured using a GHz oscilloscope. The 11.6 ns rise and 7.0 ns fall times of the input pulse are not significantly broadened, showing that ns-scale transients can be effectively delivered and the expected sub-ns temporal resolution of the electron probe is provided. Granted, some $\approx$45 MHz ringing is introduced at the edges of the pulse, but this is reproduced by a circuit model of the setup\cite{SI}, and study of this model suggests that using lower inductance components can mitigate this behavior.

In addition, the deflection magnitude is consistent with that expected for a 20 V pulse applied to this electrode geometry. The deflection along the x direction, $\theta_x$, is given by
\begin{equation}
    \theta_x = -\frac{\lambda\sigma_e}{2\pi}\int{E_x(z)}dz
\end{equation}
where $\lambda$ is the relativistic electron wavelength, $\sigma_e$ is the relativistic electron interaction parameter\cite{Kirkland_2010}, and $E_x(z)$ is the x component of the electric field at position z along the electron beam path. We performed 3D electrostatic simulations to calculate the electric field distribution in this geometry\cite{SI}. The predicted deflection is about 96 \textmu rad at 20 V (4.81 \textmu rad V$^{-1}$) at the center of the hole, which agrees with the measured 95 \textmu rad amplitude. This confirms that the local electric field measurement is accurate and that the voltage pulses are not significantly attenuated. We have also verified a linear relationship between applied field and measured deflection using DC bias from -40 to +40 V\cite{SI}.

In the future, chips could be designed to impart modified field distributions for condensed matter and other experiments. Longer electrodes with uniform separation can provide a more laterally uniform field. The contribution of stray fields can be reduced by using a narrower gap, though accurate models should still generally include both internal and stray fields especially for materials thinned to electron transparency\cite{gatel2022operando}. Finally, there is opportunity to increase the magnitude of applied transient fields. This chip is computed to deliver a peak field of about 2.8 MV m$^{-1}$ at 20 V\cite{SI}, whereas it could potentially exceed 100 MV m$^{-1}$ if the gap is reduced below 200 nm, accessing regimes of strong electric field-driven physics such as dielectric breakdown\cite{wu2004nspulsedb} and field-induced ferroelectricity\cite{li2019thzsto}.

\section{\label{sec:TaS2}Correlating field and material dynamics}

\begin{figure}
    \centering
    \includegraphics[width=3.6in]{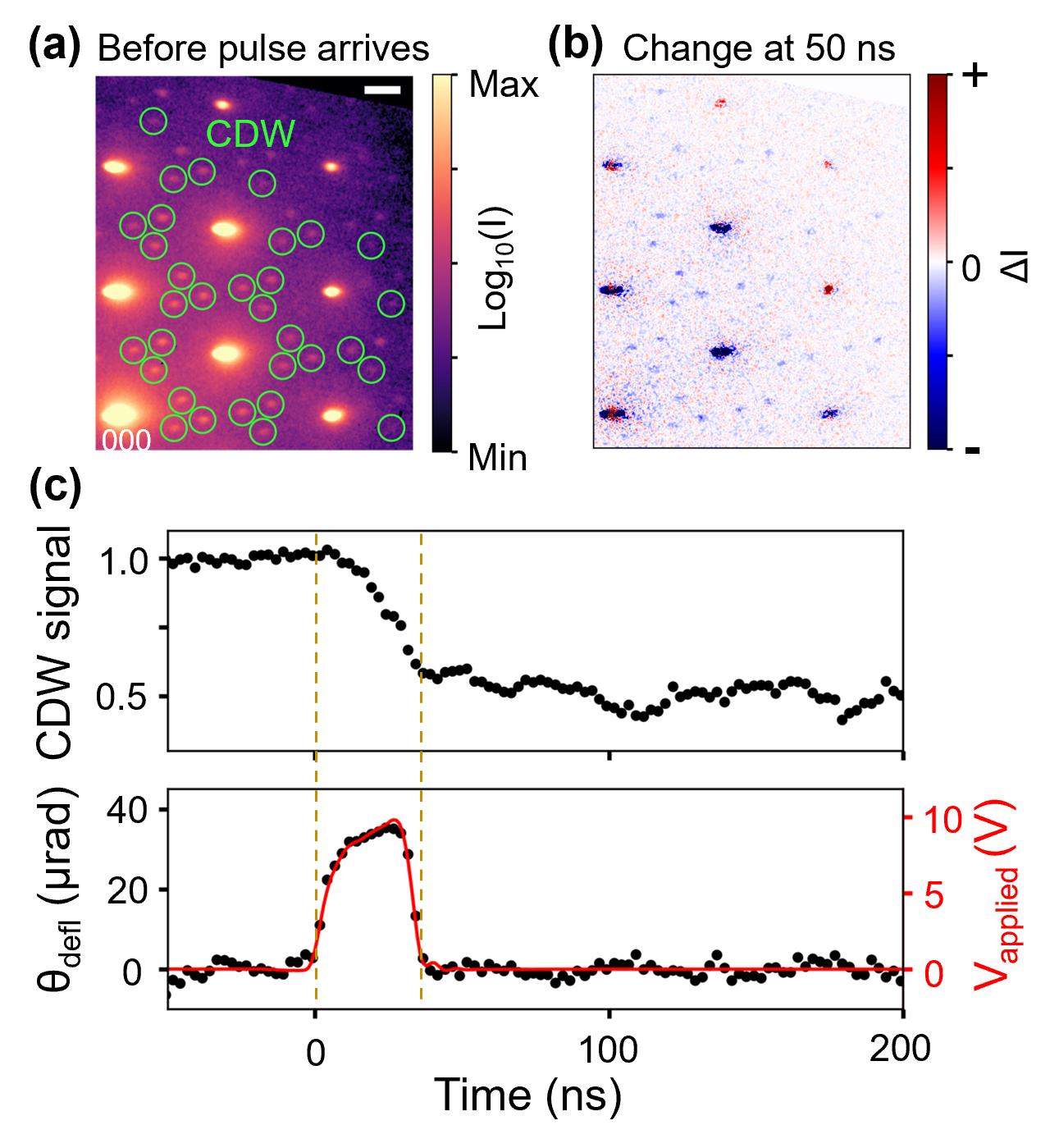}
    \caption{Nanosecond electron diffraction of (partial) phase transition in \ce{TaS2} during a 20 ns voltage pulse. a) Diffraction pattern before the pulse arrives. Superlattice peaks associated with the nearly-commensurate charge density wave (CDW) are circled in green. Scale bar indicates 1 nm$^{-1}$. b) Change in the pattern after the pulse at t = 50 ns. c) Correlated time-resolved intensity of the CDW peaks and electron probe deflection (black dots). Yellow dashed lines indicate start and end of the voltage pulse. The input voltage pulse shape is superimposed in red.}
    \label{fig:diffraction}
\end{figure}

We next demonstrate that this same diffraction configuration can simultaneously record the time-resolved field and crystal structure change in materials. 1T-\ce{TaS2} hosts a well-studied nearly-commensurate charge density wave (CDW) state at room temperature, and it is known that a phase transition to an incommensurate CDW or a normal state can be achieved using steady-state heating or bias\cite{geremew2019bias,liu2016charge} It is also known that this transition can occur within sub-picosecond timescales if triggered with optical pulses\cite{dwaynemiller2010TaS2,ruan2015TaS2,ropers2021TaS2UEM}. Here, we resolve a nanosecond-scale transition in this material during electrical pulsing. 

For this experiment, the heat/bias chip was modified using Ga focused ion beam milling to create a narrower, 1 \textmu m vacuum gap (13.5 \textmu m in length) with 1 \textmu m wide, 120 nm thick Pt electrodes deposited on either side using ion beam-induced deposition. A 100 nm thick flake of \ce{TaS2} was then exfoliated and transferred across these electrodes using polydimethylsiloxane (PDMS) stamps\cite{Castellanos_Gomez_2014,SI}. After installing in the microscope, the sample area was illuminated with the electron beam for about an hour to achieve stable electrical contact between the sample and electrodes: we attribute this to formation of a conducting surface carbon layer contacting the Pt and \ce{TaS2}. This was done until the measured resistance stabilized at 3 k$\Omega$. The initial selected area diffraction pattern in Figure~\ref{fig:diffraction}a shows that the crystal is oriented along [001], and superlattice peaks associated with the nearly-commensurate CDW state are visible at room temperature. 

We applied 20 ns long, 10 V pulses at 10 kHz and recorded the selected area electron diffraction pattern from a 1.2 \textmu m region for pump-probe delays ranging from -50 to 200 ns. Three sweeps were recorded by incrementally increasing, decreasing, and again increasing the delay. At this condition, about half of the CDW peak signal is suppressed, indicating partial phase transition throughout the probed region. This can be seen in the change in the diffraction pattern from before pulse arrival to afterwards (t = 50 ns), shown in Figure~\ref{fig:diffraction}b: While the primary lattice peaks show a complex mixture of intensity changes, all of the superlattice peaks show a strong decrease, consistent with CDW suppression. 

From this dataset, we extract and compare two time-dependent signals: the total CDW signal, obtained by summing the signal within the green circles after subtracting the local diffuse background, and the beam deflection, obtained by the center-of-mass shift of the central beam as done for the bare electrode measurements. These are shown together in Figure~\ref{fig:diffraction}c. Again, the measured deflection corresponds well to the input pulse shape, showing that such a short pulse is delivered to the material with high fidelity and allowing direct correlation between the transient field and the crystal structure. The CDW state is gradually suppressed over the duration of the pulse, and shows little change after the pulse ends, entering a period of slow recovery before the next pulse arrives 100 \textmu s later. Despite excitation with more than 7 $\times$ 10$^8$ pulses over the course of the measurement, the integrity of the sample was maintained, as evidenced by consistent dynamics among the three individual sweeps and preserved crystallinity\cite{SI}. Determining the roles of Joule heating, field, and current is beyond the scope of this work but warrants more detailed future study.

\begin{figure}
    \centering
    \includegraphics[width=3.5in]{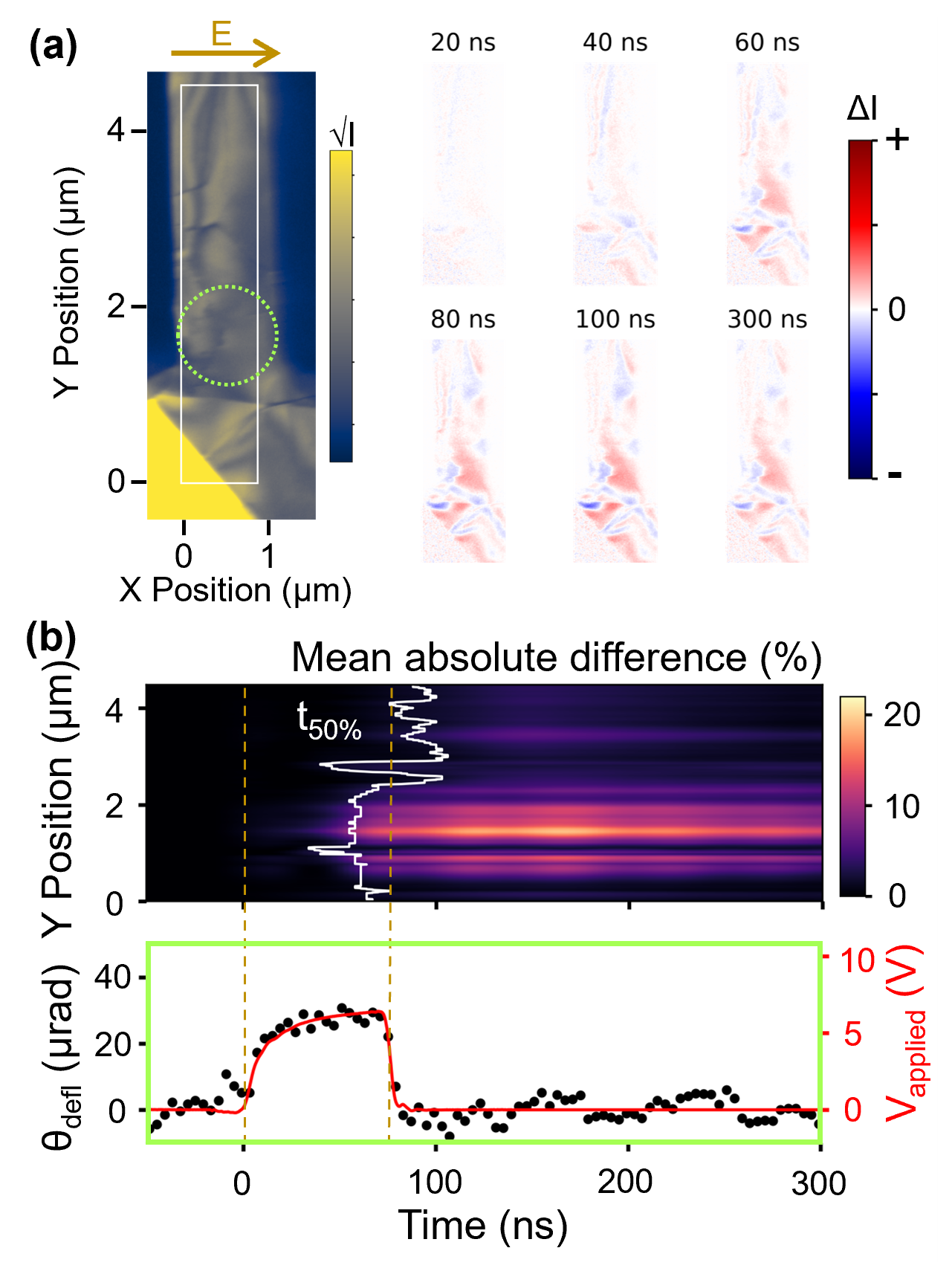}
    \caption{Nanosecond electron imaging of mesoscopic bending during a 65 ns voltage pulse. a) (Left) Initial real-space bright-field TEM image. Region for mean absolute difference analysis is indicated by the white box, while the probed area for time-resolved field measurement is indicated by the green circle. (Right) Series of difference images with respect to the initial image at selected time points. b) Time-resolved mean absolute difference for varying (y) position along the channel and corresponding electron probe deflection (black dots). The mean is computed along the x direction. The time at which the change first exceeds 50\% of the max value ($t_{50\%})$ is indicated by a white line. In the lower plot, gold dashed lines indicate start and end of the voltage pulse. The input pulse shape as measured separately using an oscilloscope is superimposed in red.}
    \label{fig:imaging}
\end{figure}

Finally, we demonstrate real-space imaging of material dynamics with nanometer-nanosecond spatiotemporal resolution. For this experiment, we used the same \ce{TaS2} flake. The initial image in Figure~\ref{fig:imaging}a shows the freestanding flake suspended across the aforementioned 1 \textmu m gap between the two deposited Pt electrodes. Notably, dark fringes known as ``bend contours" appear throughout the flake: these arise from ripples, which modulate the image contrast via the changing diffraction condition with respect to crystal orientation\cite{williams2009tem}. As such, movement of these features can be used to track local bending of the flake. The smallest visible features here are about 45 nm wide, showing nanoscale spatial resolution, though we note this was not fully optimized in these experiments and single-digit nm resolution can be achieved.

We recorded time-resolved bright-field images during 65 ns long, 6 V pulses at 10 kHz: these longer, lower amplitude pulses were used to study the response without inducing phase transition, ie. there was no change in the CDW signal. A forward sweep and reverse sweep of the pump-probe delay were recorded in succession. Irreversible changes over the long measurement time (e.g. instrument drift, accumulated pulse effects) were extracted by computing the difference image between the frames recorded at t=0 for each sweep. This change was then linearly interpolated across the measured time points and subtracted from the individual sweeps to retrieve just the reversible pulse-induced changes, at which point the two sweeps were finally averaged together\cite{SI}. 

Resulting difference images at varying time delays relative to the image at t = 0 are shown in Figure~\ref{fig:imaging}a, illustrating the contrast changes as the flake buckles in response to the electrical pulse. Notably, there is larger contrast change in the lower region compared to the upper region. We visualize the variation in dynamics along the channel length by plotting the time-resolved mean absolute difference (MAD) averaged across the gap (i.e. along the x direction). It is shown Figure~\ref{fig:imaging}b alongside the time-resolved beam deflection, recorded via a separate measurement in the indicated selected area. Again, stronger contrast changes are observed in the lower region. However, the response of the upper region is also delayed with respect to the lower. This can be seen in the contour line marking the time at which the change first exceeds 50\% of the max value ($t_{50\%})$: Most of the change in the lower region occurs during the higher amplitude portion of the pulse, while in the upper region, the changes are smaller and mostly occur after the pulse ends. A possible explanation based on a Joule heating mechanism is that transient current flows predominantly through the lower region of the device to generate initial heating and local strain there, and the upper region responds later as the induced heat equilibrates across the flake. Regardless, such spatial heterogeneity would be missed by global measurement techniques, and can be important to account for the overall material behavior. 

Taken together, the measurements on \ce{TaS2} reveal a few characteristics of dynamic behaviors during electrical pulsing. First, CDW melting can be repeatably triggered using 20 ns electrical pulses for millions of cycles without damaging the CDW state or the flake. Second, the melting occurs largely within the pulse duration. Thirdly, strain develops over the duration of the pulse even below the phase transition threshold which, in a freestanding flake, leads to bending. The speed and reversibility bode well for electronics applications. It will be important to understand how the phase transition as well as thermal and mechanical responses are distributed laterally and, looking forward to application, how they are influenced when scaling to device-like geometries. In addition, future studies should investigate the roles of Joule heating, electric field, and electric current in the observed dynamics, as well as search for novel nonequilibrium dynamics which may emerge during shorter duration and higher field pulse excitation in this and other strongly correlated materials.

\section{\label{sec:conclusions}Conclusions}
Altogether, this work demonstrates a technique that can directly measure and correlate local transient electric fields with resulting structural dynamics in materials, from the atomic-scale crystal structure to nanoscopic features. The capability to apply short, high-amplitude pulses and visualize transient dynamics within provides access to physical regimes that cannot be achieved via steady-state biasing due to breakdown or electrical heating. We anticipate that this approach will enable breakthroughs in understanding field- and current-driven phenomena in condensed matter physics such as electrical control of correlated electron phases and polarization states, dielectric breakdown mechanisms, modulation of electro-optic response, and much more. We also anticipate that this technique will find use in applied physics research, such as to visualize microelectromechanical system dynamics and failure modes as well as switching mechanisms in materials and devices for next-generation computing such as selectors and non-volatile synaptic memory.

This technique was implemented mostly using commercially available equipment, and could be implemented in many ultrafast electron microscopy and diffraction setups. In the future, dark-field imaging can be applied to visualize domain formation and growth during electrically driven phase transitions and polarization switching in crystals\cite{ropers2021TaS2UEM,nelson2011FEvortex}. It can also be readily extended to STEM and TEM spectroscopy to examine chemical, bonding, plasmonic, and phononic dynamics. Further extension to Lorentz microscopy and electron holography would allow imaging of nanoscopic magnetic and electric field texture dynamics, such as domain and vortex motion, growth, and interaction. Nanosecond electron microscopy during electrical pulsing presents a broad frontier with a host of nonequilibrium phenomena on the horizon.

\begin{acknowledgments}

This material is based upon work supported by the U.S. Department of Energy, Office of Science, for support of microelectronics research, under contract number DE-AC0206CH11357. Work performed at the Center for Nanoscale Materials, a U.S. Department of Energy Office of Science User Facility, was supported by the U.S. DOE, Office of Basic Energy Sciences, under Contract No. DE-AC02-06CH11357. We acknowledge Yue Li and Xuedan Ma at Argonne National Laboratory for their assistance with transferring the \ce{TaS2} flake.

The submitted manuscript has been created by UChicago Argonne, LLC, Operator of Argonne National Laboratory (“Argonne”). Argonne, a U.S. Department of Energy Office of Science laboratory, is operated under Contract No. DE-AC02-06CH11357. The U.S. Government retains for itself, and others acting on its behalf, a paid-up nonexclusive, irrevocable worldwide license in said article to reproduce, prepare derivative works, distribute copies to the public, and perform publicly and display publicly, by or on behalf of the Government. The Department of Energy will provide public access to these results of federally sponsored research in accordance with the DOE Public Access Plan. \url{http://energy.gov/downloads/doe-public-access-plan}

T.E.G. and D.B.D. contributed equally to this work.

\end{acknowledgments}


\nocite{*}

\bibliography{main}

%
%
%
%
%
%
%


\title{Supplemental Material \\
Nanosecond electron imaging of transient electric fields and material response}

\date{\today}


\onecolumngrid

\newpage

\renewcommand{\partname}{}
\renewcommand{\thesection}{\arabic{section}}
\renewcommand{\thetable}{\arabic{table}}   
\renewcommand{\tablename}{\textbf{Supplemental Table}}
\renewcommand{\theequation}{S\arabic{section}.\arabic{equation}}  
\renewcommand{\figurename}{\textbf{Supplemental Figure}}
\graphicspath{{figures/}} 
\setcounter{figure}{0} 
\setcounter{page}{0}

\maketitle

\onecolumngrid

\newpage


\section{Forward and reverse deflection curves}

To ensure that features observed in the time-resolved deflection measurement were indeed due to the changes in the pump-probe delay rather than instrument drifts over the course of the experiment, we performed the measurement twice: once by incrementally increasing the pump-probe delay, and the other by decreasing it. The resulting curves are shown in Supplemental Figure \ref{fig:Sweeps}. The width of the pulse, sharpness of the edges, and the observed ringing are consistent between both measurements, showing that this measurement is robust and confirming that these features represent the true time-resolved behavior of the electric field between the electrodes.

\begin{figure}[h!]
    \centering
    \includegraphics[width=11cm]{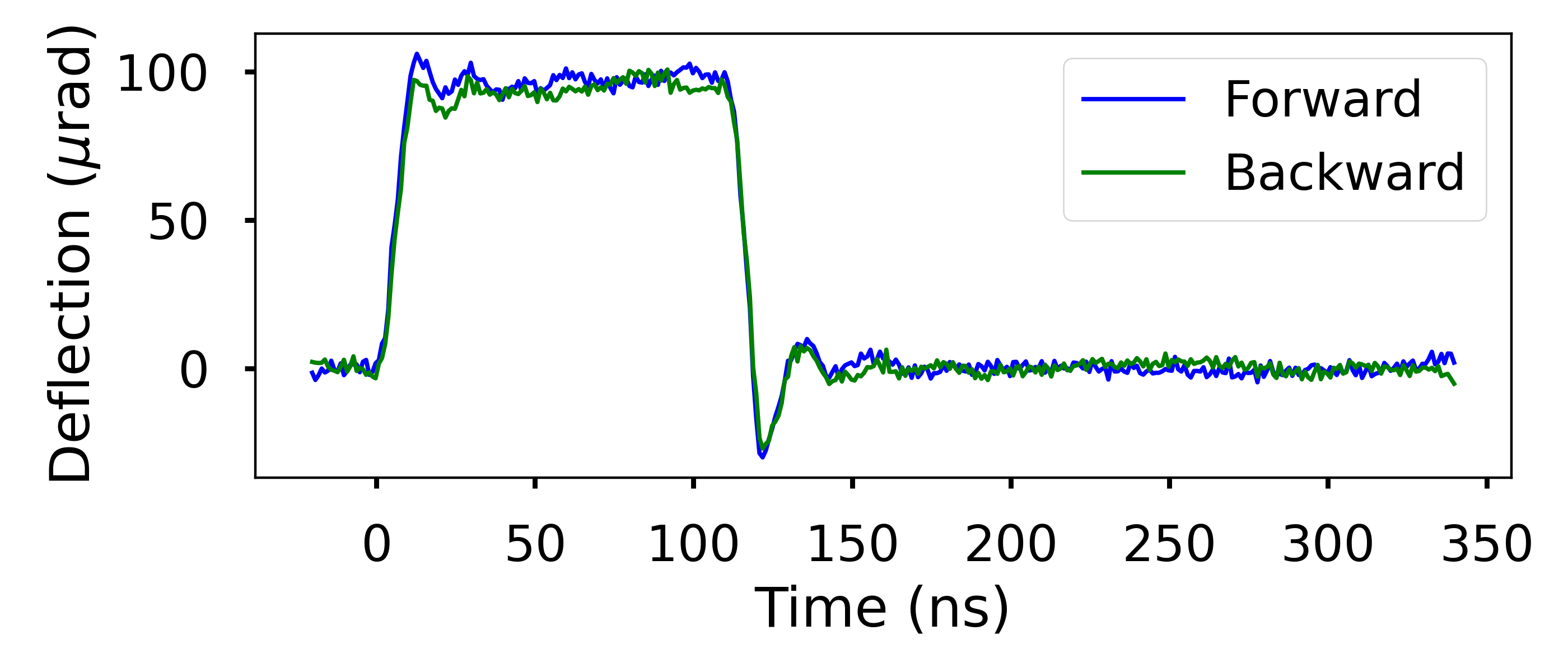}
    \caption{The two sequential time-resolved electron beam deflection curves averaged to produce the data in Figure 1 of the main manuscript. One was recorded by stepping the delay between the voltage pulse and electron probe in ascending order (Forward, blue line) while the other was in descending order (Backward, green line).}
    \label{fig:Sweeps}
\end{figure}

\newpage

\section{Dynamic circuit model}

To understand the differences between the input pulse shape and that measured between the electrodes, we constructed a circuit model using Texas Instruments TINA-TI software to simulate the time-dependent response. The input pulse shape as measured by the GHz oscilloscope is used as the input to the simulation. Including the pulse delivery cables in the model is important due to the high frequency nature of the features.  Here, a 12 inch RG58C/U BNC cable is modeled as 8 individual inductor and capacitor components to better represent its continuous nature. Values for the components in the "DENS TEM holder" and BNC to Lemo Cable sections of the model were provided by the vendor, DENS Solutions, except for the inductance which was estimated based on cable length.  The model is detailed in Supplemental Figure  \ref{fig:Circuit_Model}. The time-dependent output is plotted in Figure 1c.  

\begin{figure}[h]
    \centering
    \includegraphics[width=17cm]{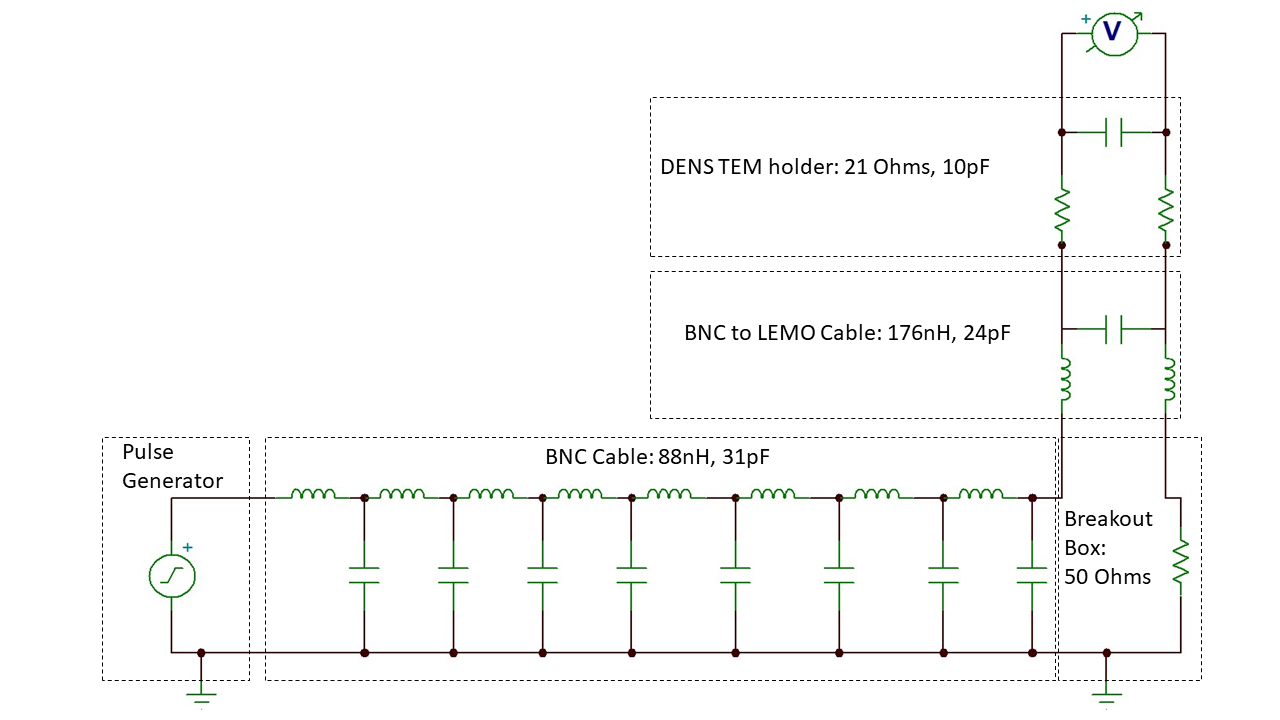}
    \caption{Transient circuit modeling for experimental setup. Circuit diagram showing the components of the setup including the voltage source, delivery cable, breakout box, connector cable and in situ biasing holder. }
    \label{fig:Circuit_Model}
\end{figure}

\newpage

\section{3D electrostatic simulation}

To estimate the integrated lateral electric field that should be experienced by the electron beam passing between the electrodes on the chip, we performed a 3-dimensional electrostatic simulation using COMSOL Multiphysics. The chip geometry is illustrated by the optical microscope image shown in Supplemental Figure \ref{fig:esSim}a. This image was used to confirm the lateral dimensions of the features that were provided by the chip manufacturer. All four gold electrodes are included in the simulation: The inner left electrode is held at ground, the inner right at 1 V, and the outer electrodes are floating. They are 180 nm thick (confirmed by AFM measurements) and the relative permittivity is set to -$\infty$ (the low-frequency limit for a metal). They are seated on a 450 nm thick silicon nitride membrane, with relative permittivity set to 7 as quoted by Norcada, a manufacturer of similar silicon nitride membranes. The electron beam passes through the 5 \textmu m gap at the end of the electrodes, which is etched completely through. The outer faces of the cell are set to a ``zero-charge" boundary condition, where the component of the dielectric displacement vector along the face normal is forced to zero. The simulation cell size and sampling density were increased until the integrated electric field values of interest converged to within 1\%. The final cell size was 100 \textmu m x 100 \textmu m x 200 \textmu m. 

The results of the simulation are shown in Supplemental Figure \ref{fig:esSim}b-e. The potential map in panel b confirms that the inner electrodes show a 1 V potential difference, while the floating outer electrodes take on a smaller constant potential to negate internal currents. The electric field component in the direction across the gap $E_x$ is mapped in the xy plane in panel c. Indeed, the electric field is maximum in the narrow gap at the end of the electrodes. As shown in the yz plane map in panel d, this electric field extends well above the thickness of the electrodes, and the electric field integrated along the electron beams path from top to bottom $\int E_x dz$ leads to the electron beam deflection. The integrated field as a function of position relative to the center of the hole is plotted in panel e: it is uniform across the gap (x direction), but along the channel (y direction) there is a gradient as the tip of the electrodes sees less field compared to the middle. 

\begin{figure}[h!]
    \centering
    \includegraphics[width=6in]{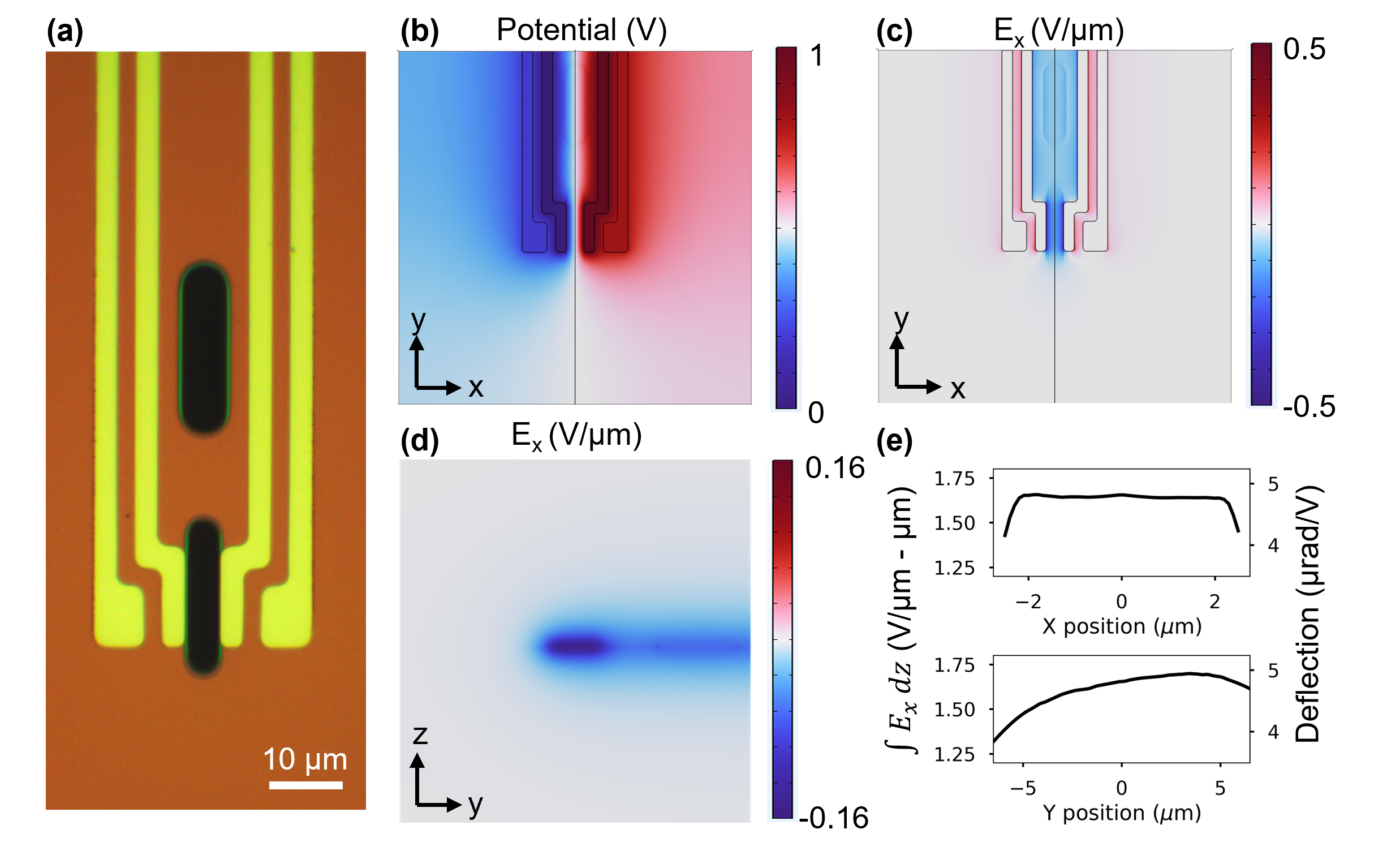}
    \caption{3D electrostatic simulation of the electric field distribution in the chip. a) Optical microscope image showing the gold electrodes on the reddish-brown silicon nitride membrane with visually black through holes. b) Lateral slice of the simulated electrostatic potential distribution taken at the middle of the electrode thickness. The edges of the electrodes and the center line are superimposed in black. c) The same, but for the x-component of the electric field ($E_x$). d) y-z plane slice taken from the center. e) Profiles of $\int E_x dz$ and corresponding electron beam deflection across the gap (top) and along the channel (bottom).}
    \label{fig:esSim}
\end{figure}

\newpage

\section{Steady-state beam deflection calibration}
At the same microscope conditions as for the time-resolved beam deflection measurements, we also recorded the deflection under varying DC bias. The results are shown in Supplemental Figure \ref{fig:DCdefl}. The behavior is linear from -40 V to 40 V. The least squares fit line shown in red retrieves a deflection per applied voltage of 4.16 \textmu rad V$^{-1}$, which in the electrostatic simulation corresponds to a beam position located towards the end of the electrodes (see Supplemental Figure \ref{fig:esSim}e).

\begin{figure}[h!]
    \centering
    \includegraphics[width=8cm]{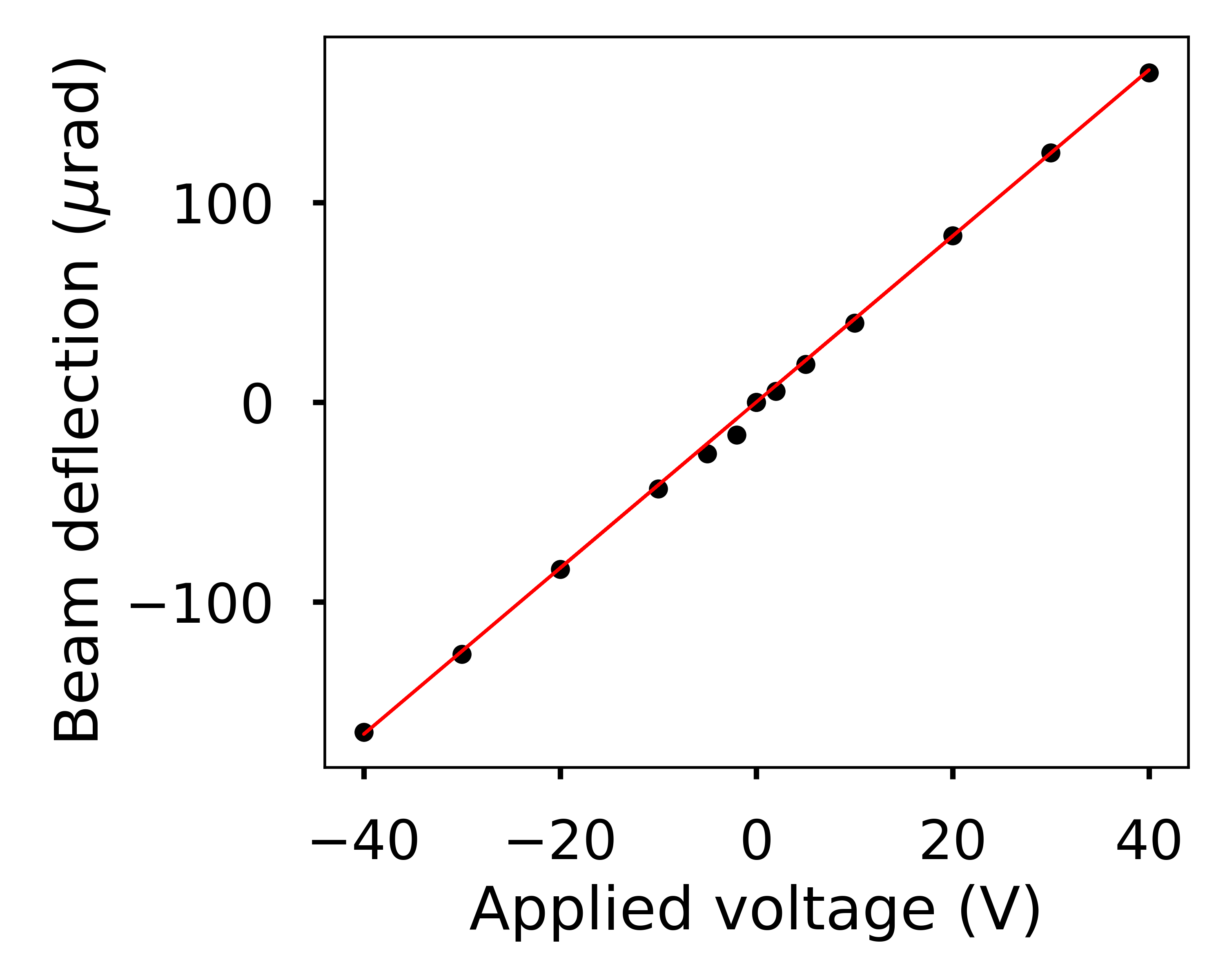}
    \caption{Electron beam deflection through the pristine chip at varying DC bias. Measured points are given by the black dots, while the linear fit is shown in red.}
    \label{fig:DCdefl}
\end{figure}

\newpage

\section{Sample preparation}
For the \ce{TaS2} experiments, the heat/bias chip was first modified using 30 kV Ga focused ion beam milling to create a narrower, 1 \textmu m vacuum gap (13.5 \textmu m in length).  Then, 1 \textmu m wide, 120 nm thick Pt electrodes were deposited on either side using ion beam-induced deposition with 30 kV Ga ions at 100 pA beam current. A 100 nm thick flake of \ce{TaS2} (thickness verified by AFM) was then exfoliated and transferred across these electrodes via PDMS stamping. The resulting sample is shown in Supplemental Figure \ref{fig:Sample}.

\begin{figure}[h!]
    \centering
    \includegraphics[width=7cm]{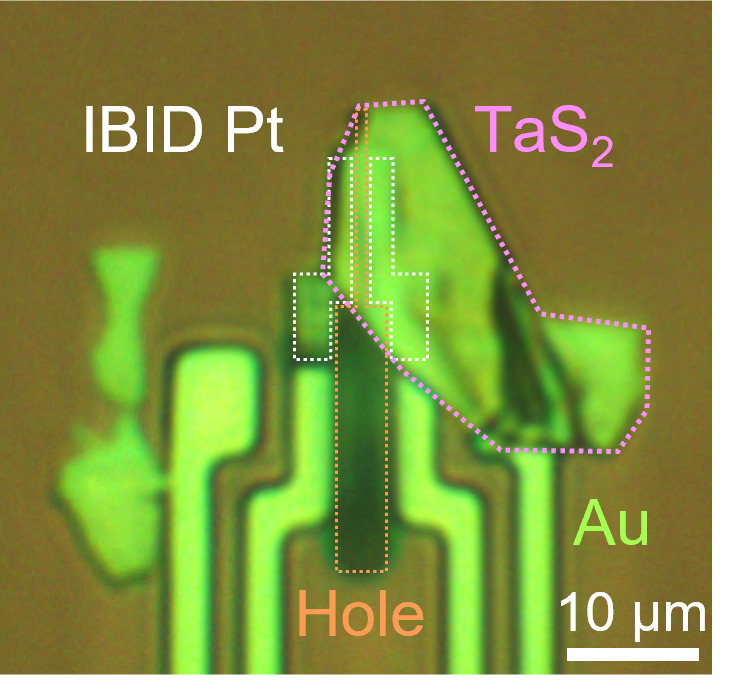}
    \caption{Light microscope image of the sample. IBID is ion beam-induced deposition.}
    \label{fig:Sample}
\end{figure}

\newpage

\section{Sample integrity}
The sample integrity was verified by comparing the average of patterns for time delays less than 0 (before pulse arrival) for the first and third sweeps, shown in Supplemental Figure \ref{fig:DPsBeforeAfter}.

\begin{figure}[h!]
    \centering
    \includegraphics[width=14cm]{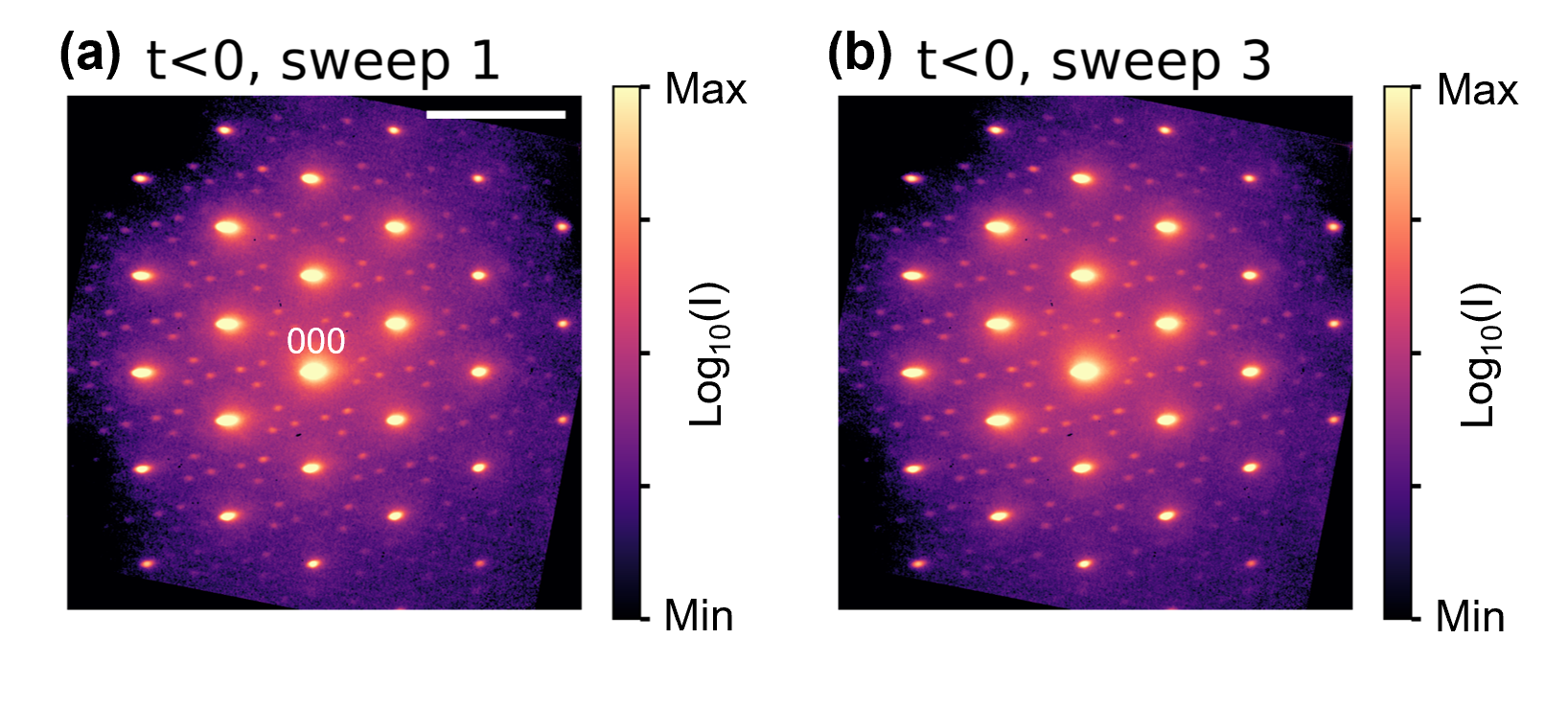}
    \caption{Sample integrity over the course of the time-resolved diffraction measurement shown in Fig 2 of the main manuscript. (a) Average of patterns recorded at t$<$0 for the first sweep. (b) Average of patterns recorded at t$<$0 for the third sweep. Scale bar = 5 nm$^{-1}$}
    \label{fig:DPsBeforeAfter}
\end{figure}

\newpage

\section{Separating reversible and irreversible dynamics in time-resolved imaging}
To separate the contrast changes which are repeatable due to the electrical pulse excitation from irreversible effects due to accumulated pulses and instrument drift over the course of the measurement, we recorded both a forward sweep stepping the pump-probe delay from -50 ns to +300 ns and then a reverse sweep from +300 ns to -50 ns. The irreversible component was computed as the difference between time zero images recorded in the forward and reverse sweeps. This was then subtracted by approximating this as a linear change over the measurement and the resulting forward and reverse sweeps were averaged to produce the data shown in Figure 3 of the main article. The individual sweeps after this subtraction as well as the irreversible component subtracted are shown in Supplemental Figure \ref{fig:ReverseSweep}. The time dependent dynamics observed in the forward and reverse sweeps show good agreement after this procedure. The extracted irreversible component, shown in Supplemental Figure \ref{fig:ReverseSweep}c, appears to be large but it should be noted that it occurs slowly over the duration of the entire measurement, so the time-dependent pulse-induced changes are still dominant and readily extracted from the data. 

\begin{figure}[h!]
    \centering
    \includegraphics[width=14cm]{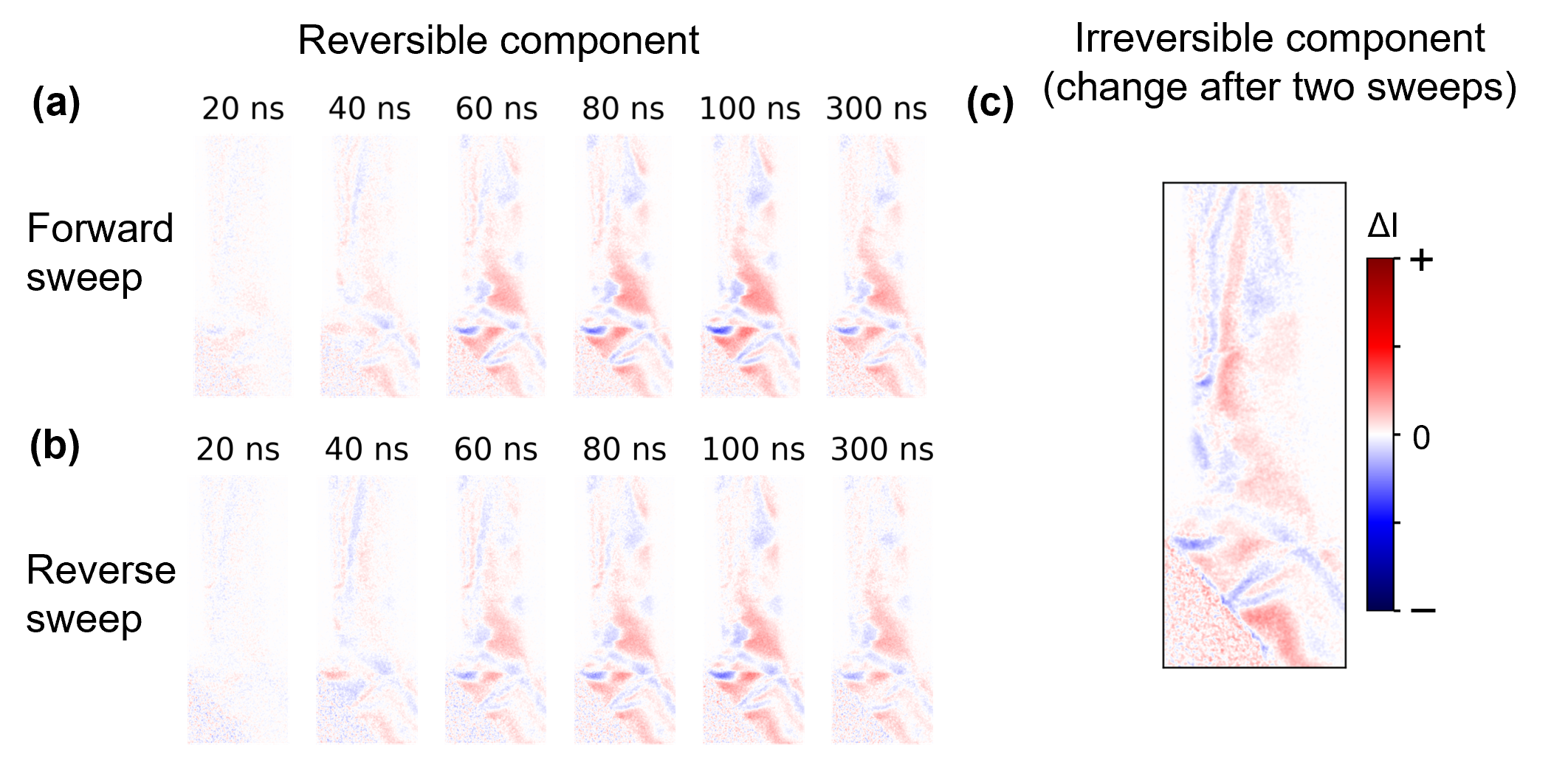}
    \caption{Separated reversible and irreversible changes (a) Forward sweep after subtracting irreversible component (b) Same but for reverse sweep. (c) Irreversible component extracted by taking the difference between time zero images recorded in the sequential forward and reverse sweeps.}
    \label{fig:ReverseSweep}
\end{figure}

%

\end{document}